\documentclass[12pt]{article}
\usepackage{amssymb}
\usepackage{graphicx}
\usepackage[cp1251]{inputenc}
\usepackage{rotating}

\begin{document}
 \noindent {\footnotesize\it Astronomy Letters, 2023, Vol. 49, No. 3, pp. 184--192}
 \newcommand{\dif}{\textrm{d}}

 \noindent
 \begin{tabular}{llllllllllllllllllllllllllllllllllllllllllllll}
 & & & & & & & & & & & & & & & & & & & & & & & & & & & & & & & & & & & & & &\\\hline\hline
 \end{tabular}

  \vskip 0.5cm
  \bigskip
 \bigskip
 \centerline{\large\bf Determination of the Spiral Pattern Speed in the Galaxy}
 \centerline{\large\bf from Three Samples of Stars}

 \bigskip
 \bigskip
  \centerline { %  DOI: 10.1134/S1063773723030027
   V. V. Bobylev\footnote [1]{vbobylev@gaoran.ru},  A. T. Bajkova}
 \bigskip
 \centerline{\small\it Pulkovo Astronomical Observatory, Russian Academy of Sciences, St. Petersburg, 196140 Russia}
 \bigskip
 \bigskip
{We invoke the estimates of the amplitudes of the velocity perturbations $f_R$ and $f_\theta$ caused by the influence of a spiral density wave that have been obtained by us previously from three stellar samples. These include Galactic masers with measured VLBI trigonometric parallaxes and proper motions, OB2 stars, and
Cepheids. From these data we have obtained new estimates of the spiral pattern speed in the Galaxy $\Omega_p:$
 $24.61\pm2.06$, $24.71\pm1.29$ and $25.98\pm1.37$~km s$^{-1}$ kpc$^{-1}$
from the samples of masers, OB2 stars, and Cepheids, respectively. The corotation radii for these three samples $R_{\rm cor}/R_0$ are $1.16\pm0.09$, $1.15\pm0.06$ and $1.09\pm0.06,$ suggesting that the corotation circle is located between the Sun and the Perseus arm segment.
 }

\bigskip
\section*{INTRODUCTION}
Studying the spiral structure of the Galaxy is of great interest. The spiral structure tracers, for example, hydrogen clouds, OB stars, or Cepheids, are well
known. Various methods of determining such parameters as the spiral pattern pitch angle $i,$ the number of spiral arms $m,$ the spiral pattern speed $\Omega_p,$ and
the position of the corotation radius $R_{\rm cor}$ have been proposed. However, there is no complete agreement between the results of various authors.

Beginning with the pioneering studies of Lin and Shu (1964), Lin et al. (1969), and Yuan (1969) devoted to the application of the linear spiral density wave theory to the analysis of real data, a huge number of scientific publications are devoted to this problem. For example, the studies of Marochnik et al. (1972), Cr\'ez\'e and Mennessier (1973), Byl and Ovenden (1978), Mishurov et al. (1979), Loktin and
Matkin (1992), Mishurov et al. (1997), Amaral and
L\'epine (1997), Rastorguev et al. (2001), Fern\'andez et al. (2001), Dias and L\'epine (2005), Junqueira et al. (2015), Dambis et al. (2015), Dias et al. (2019), Castro-Ginard et al. (2021), and Joshi and Malhotra (2023) can be noted.

The two-armed model of a spiral pattern with $m=2$ and $i\sim -6^\circ$ has often been used previously. In recent years, there has been more inclination toward
the four-armed model with $m=4$ and $i\sim -12^\circ$. A large body of evidence precisely for the four-armed global spiral pattern was collected in the reviews by
Vall\'ee (1995, 2002, 2008, 2017a), although the case
in point is a global spiral pattern with a constant pitch
angle, the same for all arms. In recent years, however,
the four-armed model with a sector structure of arms (Reid et al. 2014, 2019), which is substantiated by the analysis of masers with highly accurate VLBI measurements of their trigonometric parallaxes, has gained in popularity.

Accurate values of the spiral pattern speed and the corotation radius are of great interest. However, the present-day estimates of these parameters lie in
a wide range. For example, it was concluded in
the review by Gerhard (2011) that the spiral pattern
speed is slightly smaller than the rotation rate of the
Galaxy at the solar distance $R_0.$ This means that the
corotation radius is slightly farther than $R_0.$ According
to tracers with ages of $10^7-10^8$ yr, the average
$\Omega_p$ is $25.2$~km s$^{-1}$ kpc$^{-1}$. However, the studies
devoted to the distribution of stellar velocities in the
solar neighborhood give a wider range of $\Omega_p:$ 17--28~km s$^{-1}$ kpc$^{-1}$. Jacques Vall\'ee regularly reviews
the parameters of the Galactic spiral structure. In one
of his latest studies he found (Vall\'ee 2017b) that, on
average, $\Omega_p$ is close to $23\pm2$~km s$^{-1}$ kpc$^{-1}$.

The position of the corotation radius is of great and, occasionally, critical importance in modeling some processes. The point is that in a rotating reference
frame the density wave moves from corotation to the Galactic center and from corotation into the outer Galaxy. As shown by Acharova et al. (2010),
the combined effect of the corotation resonance and
turbulent diffusion is responsible for the formation of
a bimodal radial distribution of iron and oxygen in the
Galactic disk. Another example: open star clusters in
the corotation region, while undergoing small radial
oscillations, are scattered upon disruption over a huge
disk space (Mishurov and Acharova 2011).

In a number of our papers (Bobylev and Bajkova 2022a, 2022b, 2022d) we found the amplitudes of the velocity perturbations $f_R$ and $f_\theta$ caused by the
influence of a spiral density wave and estimated the
angular velocity of Galactic rotation. These parameters
were found from three samples: from Galactic
masers with measured VLBI trigonometric parallaxes
and proper motions, OB2 stars, and Cepheids. To
determine $f_R$ and $f_\theta$, we applied a spectral analysis of
the residual stellar velocities. The goal of this paper is to estimate the spiral pattern speed $\Omega_p$ in the Galaxy and the position of the corotation radius $R_{\rm cor}$ based on these data.

 \section*{METHOD}\label{method}
The position of a star in a logarithmic spiral wave can be written as
 \begin{equation}
 R=R_0 e^{(\theta-\theta_0)\tan i},
 \label{spiral-1}
 \end{equation}
where $R$ is the Galactocentric distance of the star;
$R_0$ is the Galactocentric distance of the Sun; $\theta$ is
the position angle of the star: $\tan\theta=y/(R_0-x)$,
where $x, y$ are the heliocentric Galactic rectangular
coordinates of the star, with the $x$ axis being directed
from the Sun to the Galactic center and the direction
of the $y$ axis coinciding with the direction of Galactic
rotation; $\theta_0$ is some arbitrarily chosen initial angle;
$i$ is the pitch angle of the spiral pattern ($i<0$ for a
winding spiral). After taking the logarithm of the left
and right parts, Eq.~(1) can be rewritten as
 \begin{equation}
  \ln\biggl(\frac{R}{R_0}\biggr)=\theta\tan i+{\rm const},
 \label{spiral-02}
 \end{equation}
According to the spiral density wave theory of Lin and Shu (1964), Eq. (2) appears as (Yuan 1969)
\begin{equation}
 \ln\biggl(\frac{R}{R_0}\biggr)=\tan i~\biggl(\theta+\frac{\chi-\chi_0}{m}-
 \Omega_p t\biggr),
 \label{spiral-44}
\end{equation}
where $\chi$ is the radial phase of the wave, $\chi_0$ is the position of the Sun in the wave, $\Omega_p$ is the spiral pattern speed, $t$ is the time, and $m$ is the number of spiral arms.

The relation (Rohlfs 1977) that follows from the linear density wave theory of Lin and Shu (1964) underlies the approach that we apply in this paper:
\begin{equation}
 \begin{array}{lll}
 \renewcommand{\arraystretch}{3.6}
 \displaystyle
\chi=m[\Omega_p-\Omega(R)] t+\ln\biggl(\frac{R}{R_0}\biggr)\cot i=  %\\
 \displaystyle
  %\qquad  =
\varkappa\nu t+\ln\biggl(\frac{R}{R_0}\biggr)\cot i,
 \label{spiral-777}
 \end{array}
\end{equation}
where $\Omega=\Omega(R)$ is the angular velocity of Galactic rotation, 
$\varkappa^2=4\Omega^2\left(1+\frac{\displaystyle R}{\displaystyle 2\Omega}
 \frac{\displaystyle d\Omega}{\displaystyle dR}\right)$ is the epicyclic frequency ($\varkappa>0$), $\nu=m(\Omega_p-\Omega)/\varkappa$ is the frequency
with which a test particle encounters the passing spiral perturbation.

The influence of a spiral density wave on the rectangular
heliocentric space velocities of a star $U$ and $V$ is periodic and, therefore, is represented as follows (Cr\'ez\'e and Mennessier 1973; Mishurov et al. 1979):
\begin{equation}
 \begin{array}{lll}
 \displaystyle
U= f_R\cos\chi,\\
 \displaystyle
V=f_\theta\sin\chi,
 \label{spiral-888}
 \end{array}
\end{equation}
where the velocity perturbation amplitudes $f_R$ and $f_\theta$ can be found from observations, for example, by solving the kinematic equations or through a spectral
analysis of the residual (freed from the peculiar solar motion and the Galactic rotation) stellar velocities. Note that both velocity perturbations $f_R$ and $f_\theta$ that we found based on our spectral analysis are positive.

On the other hand, $f_R$ and $f_\theta$ have the following form:
 \begin{equation}
 \renewcommand{\arraystretch}{2.0}
       f_R= {k A\over \varkappa}{\nu \over {1-\nu^2}} \Im^{(1)}_\nu(x),\\
       \qquad
 \label{Factors-1}
  f_\theta= -{k A\over 2\Omega} {1 \over {1-\nu^2}} \Im^{(2)}_\nu(x),
 \label{Factors-2}
 \end{equation}
where $A$ is the amplitude of the spiral wave potential, $k=m\cot (i)/R$ is the radial wave number, $\Im^{(1)}_\nu(x)$ and $\Im^{(2)}_\nu(x)$ are the reduction factors,
\begin{equation}
 \begin{array}{lll}
 \renewcommand{\arraystretch}{3.8}
 \displaystyle
  \Im^{(1)}_\nu(x)={{1-\nu^2}\over x} \biggl[1-{\nu\pi\over \sin(\nu\pi)}
    {1\over 2\pi}\int_{-\pi}^{+\pi} e^{-x(1+\cos(s))} \cos(\nu s) ds \biggr],
 \label{spiral-77}
 \end{array}
\end{equation}
\begin{equation}
 \begin{array}{lll}
 \displaystyle
  \Im^{(2)}_\nu(x)=(\nu^2-1) {\nu\pi\over \sin(\nu\pi)}
    {\partial\over\partial x}
  \Biggl[{1\over 2\pi}\int_{-\pi}^{+\pi} e^{-x(1+\cos(s))} \cos(\nu s) ds\Biggr],
 \label{spiral-99}
 \end{array}
\end{equation}
which are functions of the coordinate $x=k^2\sigma^2_R/\varkappa^2,$
where $\sigma_R$ is the root-mean-square (rms) stellar radial velocity dispersion. Relations (6)--(9) allows $\Omega_p$ to be determined after substituting the parameters ($f_R, f_\theta, \Omega_0, \Omega'_0,\sigma_R$) derived from observations.

We estimate the amplitude of the spiral wave potential
based on the well-known relation (Fern\'andez et al. 2008)
 \begin{equation}
 A=\frac{(R_0\Omega_0)^2 f_{r0} \tan i}{m},
 \label{f-r0}
 \end{equation}
where we take the ratio of the radial component of the gravitational force corresponding to the spiral arms to the total gravitational force of the Galaxy, $f_{r0},$ to be $0.04\pm0.01$ (Bobylev and Bajkova 2012). In this
paper we use the four-armed Galactic spiral pattern
($m=4$) with the pitch angle $i=-12.5^\circ$. We take
$R_0$ to be $8.1\pm0.1$ kpc, according to the review by
Bobylev and Bajkova (2021), where it was deduced as a weighted mean of a large number of present-day individual estimates.

Given the ratio $\Im^{(2)}_\nu/\Im^{(1)}_\nu$, the value of $\Omega_p$ found
can be checked according to the expression derived
from (6) and (7) for $R=R_0$ and $\Omega=\Omega_0$:
\begin{equation}
 \renewcommand{\arraystretch}{2.0}
 \Omega_p-\Omega_0=-\frac{\Im^{(2)}_\nu}{\Im^{(1)}_\nu}~
                    \frac{f_R}{f_\theta}~
 \frac{2\Omega_0}{m}\biggl(1+\frac{R_0\Omega'_0}{2\Omega_0}\biggr).
 \label{OB-1}
\end{equation}

 \section*{RESULTS}
Table 1 summarizes the results of our determination of the spiral pattern speed $\Omega_p$ and the corotation radius $R_{\rm cor}$ from three samples of young objects. These include masers, OB2 stars, and Cepheids.

We calculated the corotation radius based on the relation derived by equating the linear rotation velocity of the Galaxy and the rotation velocity of the spiral
pattern found:
\begin{equation}
R_{\rm cor}=R_0+(\Omega_p-\Omega_0)/\Omega'_0 ~.
 \label{R-cor}
\end{equation}

 \subsubsection*{Masers}
In Bobylev and Bajkova (2022b) we analyzed a
sample of masers and radio stars with measured
VLBI trigonometric parallaxes; only objects with
parallax measurement errors less than 10\% were
considered.

The catalogues by Reid et al. (2019) and Hirota
et al. (2020) served as the main sources of data.
Data on 199 masers were included in the list by
Reid et al. (2019). The VLBI observations were
carried out at several radio frequencies within the
BeSSeL (the Bar and Spiral Structure Legacy Survey \footnote{http://bessel.vlbi-astrometry.org})
project. Hirota et al. (2020) presented a catalog
of 99 maser sources observed exclusively at 22 GHz
within the VERA (VLBI Exploration of Radio Astrometry \footnote{http://veraserver.mtk.nao.ac.jp})
program.

%%%%%%%%%%%%%%%%%%%%%%%%%%%%%%%%%%%%%%%%%%%%%%%%%%%%%%%%%%%%% t-Oort
  \begin{table}[t]
  \caption[]{\small
Estimates of the spiral pattern parameters $\Omega_p$ and $R_{\rm cor}$
 }
  \begin{center}  \label{Table-1}    \small
  \begin{tabular}{|l|c|c|c|c|c|c|}\hline
   Parameters                    & Masers   & OB2 stars & Cepheids \\
   \hline
 $N_\star$                       & 150       & 1812       &     363 \\
 $\Omega_0,$~km s$^{-1}$ kpc$^{-1}$            &$ 30.18\pm0.38$  &$ 29.71\pm0.06$  & $ 28.87\pm0.23$   \\
 $\Omega^{'}_0,$~km s$^{-1}$ kpc$^{-2}$  &$-4.368\pm0.077$ &$-4.014\pm0.018$ & $-3.894\pm0.063$ \\
 $f_R,$~km s$^{-1}$                    & $8.1\pm1.4$ & $4.8\pm0.7$ & $5.5\pm2.0$ \\
 $f_\theta,$~km s$^{-1}$               & $6.1\pm1.7$ & $4.1\pm0.9$ & $7.1\pm2.0$ \\
 $\sigma_R,$~km s$^{-1}$               &          12 &        13.4 &          15 \\
         Source            &         (1) &         (2) &         (3) \\
 $\Omega_p,$~km s$^{-1}$ kpc$^{-1}$ & $24.61\pm2.06$ & $24.71\pm1.29$ & $25.98\pm1.37$ \\
 $R_{\rm cor},$~kpc             & $ 9.37\pm0.78$ & $ 9.34\pm0.49$ & $ 8.84\pm0.47$ \\
 $R_{\rm cor}/R_0$              & $ 1.16\pm0.09$ & $ 1.15\pm0.06$ & $ 1.09\pm0.06$ \\
 $\Im^{(2)}_\nu/\Im^{(1)}_\nu$  &          0.672 &          0.635 & 0.569 \\
 \hline
 \end{tabular}\end{center}
 {\small
 $N_\star$ is the number of stars used; (1) Bobylev and Bajkova (2022b); (2) Bobylev and Bajkova (2022a, 2022c); (3) Bobylev and Bajkova (2022d).
 }
 \end{table}
%%%%%%%%%%%%%%%%%%%%%%% t1

The distribution of the masers and radio stars used in projection onto the Galactic $XY$ plane is given in Fig.~1. The coordinate system in which the $X$ axis is directed from the Galactic center to the Sun and the direction of the $Y$ axis coincides with the direction of Galactic rotation is used in this figure. The four-armed spiral pattern with the pitch angle $i=-13^\circ$ from Bobylev and Bajkova (2014) is given; here, it was constructed with $R_0=8.1$~kpc, the Roman numerals number the following four spiral arms: Scutum (I), Carina-Sagittarius (II), Perseus (III), and the Outer Arm (IV).

The Local Arm ($\sim$70 sources) is well represented in Fig.~1. Nevertheless, a concentration of stars to the Perseus, Carina-Sagittarius, and Scutum arm
segments is seen.

The angular velocity of Galactic rotation and its two derivatives were estimated using 150 masers from the Galactic region $R>4$~kpc, while the velocity
perturbations $f_R$ and $f_\theta$ were estimated by applying
a spectral analysis of the residual velocities of masers
within 5 kpc of the Sun. Based on this sample of masers, we calculated the rms radial velocity dispersion to be $\sigma_R=12$ km s$^{-1}$.

 \subsubsection*{OB2 stars}
The angular velocity of Galactic rotation and its derivatives were estimated in Bobylev and Bajkova (2022a) by analyzing the proper motions of 9750 OB2 stars. The sample of OB2 stars from Xu et al. (2021) with proper motions and trigonometric
parallaxes from the Gaia EDR3 catalogue was used for this purpose.

Based on this sample, we also found the principal axes of the ellipsoid of residual velocity dispersions for OB2 stars, $\sigma_1,\sigma_2,\sigma_3)=(11.79,9.66,7.21)\pm(0.06,0.05,0.04)$ km s$^{-1}$, and showed that the first axis of this ellipsoid deviates from the direction to the
Galactic center by a small angle of about $12^\circ$. Thus, $\sigma_1$ can be used as the radial velocity dispersion $\sigma_R$ in relations (6)--(9).

The kinematics of the total space velocities of OB2 stars was studied in Bobylev and Bajkova (2022c). These include 1812 stars with measured
line-of-sight velocities, proper motions, and trigonometric
parallaxes. The distribution of these OB2 stars
in projection onto the Galactic $XY$ plane is given
in Fig. 2. We determined the velocity perturbation
amplitudes $f_R$ and $f_\theta$ from them by applying a spectral analysis of the residual velocities. As can be seen from the figure, the Local Arm as well as the Carina-Sagittarius and Perseus arms are well represented.

A direct calculation of the radial velocity dispersion based on this sample of OB2 stars gave $\sigma_R=13.4$~km s$^{-1}$. Thus, the line-of-sight velocity errors
increased $\sigma_R$ compared to $\sigma_1=11.79$~km s$^{-1}$ obtained
by analyzing only the stellar proper motions.

%%%%%%%%%%%%%%%%%%%%%%%% FIG.1:
\begin{figure}[t]
{ \begin{center}
  \includegraphics[width=0.6\textwidth]{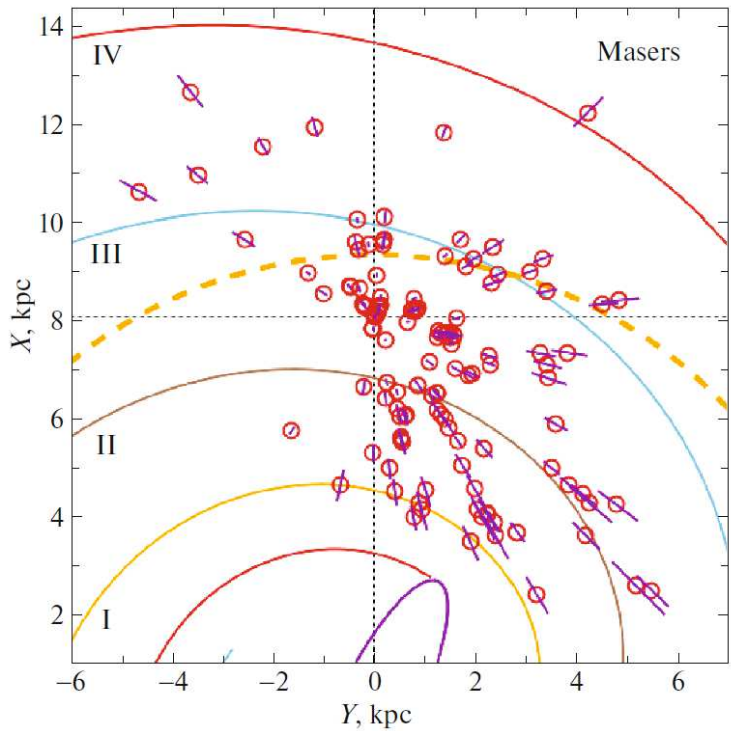}
  \caption{
Distribution of 150 masers and radio stars with trigonometric parallax errors less than 10\% in projection onto the Galactic $XY$ plane, the distance errors bars are given for each star, the four-armed spiral pattern with the pitch angle $i=-13^\circ$
is shown, the central Galactic bar is marked, the thick dashed line indicates the corotation circle found.
  }
 \label{f-masers-XY}
\end{center}}
\end{figure}
%%%%%%%%%%%%%%%%%%%% f1
%%%%%%%%%%%%%%%%%%%%%%%% FIG.2:
\begin{figure}[t]
{ \begin{center}
  \includegraphics[width=0.96\textwidth]{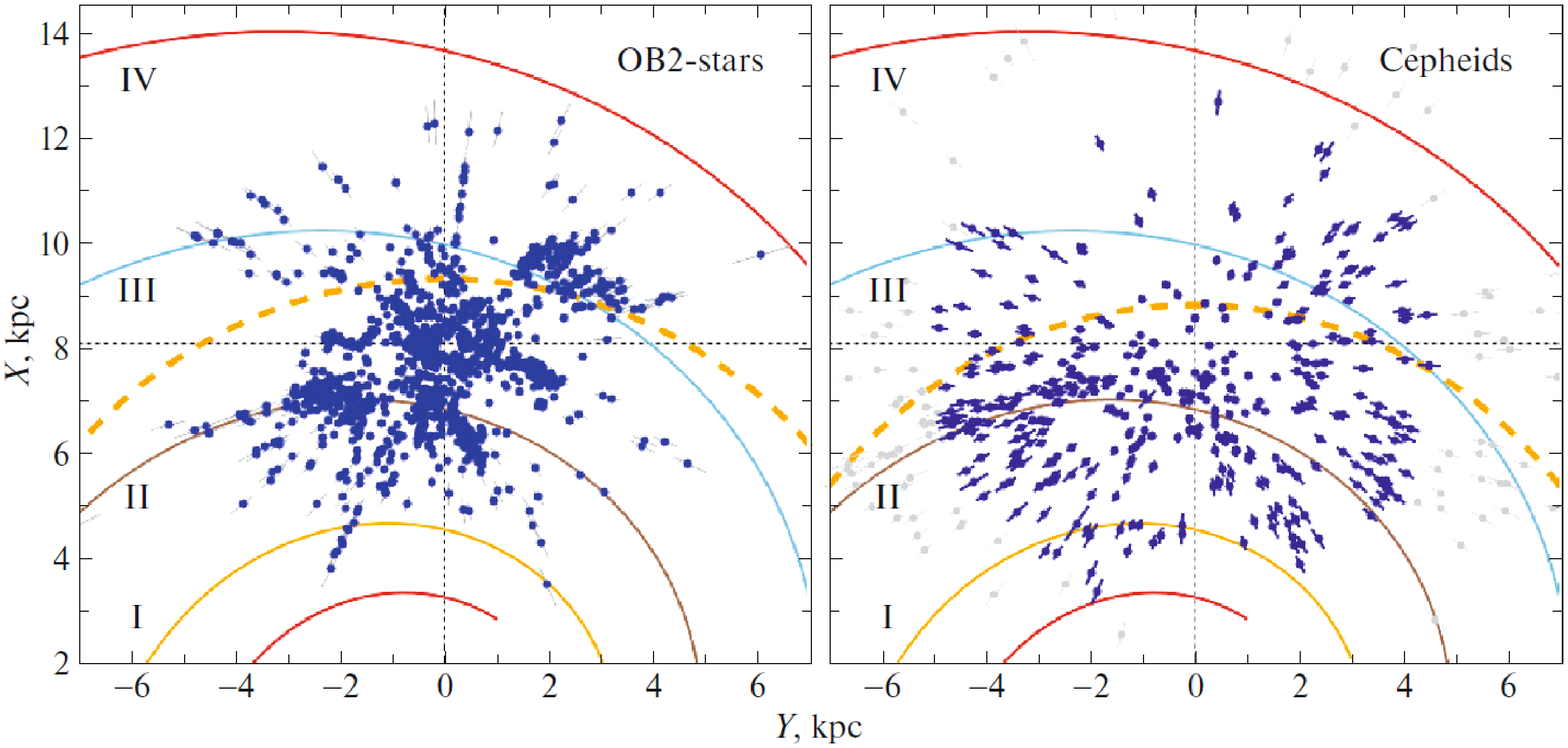}
  \caption{
Distribution of OB2 stars (left) and Cepheids (right) in projection onto the Galactic $XY$ plane, the error bars that correspond to the declared distance errors in the corresponding catalogue are given, the four-armed spiral pattern with the
pitch angle $i=-13^\circ$ is shown, the thick dashed lines indicate the corresponding corotation circles. 
  }
 \label{f-OB-Ceph-XY}
\end{center}}
\end{figure}
%%%%%%%%%%%%%%%%%%%% f2

 \subsubsection*{Cepheids}
The angular velocity of Galactic rotation and its
derivatives were estimated from a sample of Cepheids
in Bobylev and Bajkova (2022d). A sample of 363
Cepheids younger than 120 Myr located no farther
than 5 kpc from the Sun was used for this purpose.
Their average age is 85 Myr. We took the proper motions
of these stars from the Gaia EDR3 catalogue.

The paper by Skowron et al. (2019), where the
distances, ages, pulsation periods, and photometric
data are given for 2431 classical Cepheids, served as
a basis for studying the sample of Cepheids. The
observations of these variable stars were performed
within the OGLE (Optical Gravitational Lensing Experiment)
program (Udalski et al. 2015). The distances
to the Cepheids were calculated based on
the calibration period--luminosity relations found by
Wang et al. (2018) from the mid-infrared light curves
of Cepheids for eight bands. These include four bands
of the WISE (Wide-field Infrared Survey Explorer,
Chen et al. 2018) catalogue, W1--W4: [3.35], [4.60],
[11.56], and [22.09] $\mu m$, and four bands of GLIMPSE
(Spitzer Galactic Legacy Infrared Mid-Plane Survey
Extraordinaire, Benjamin et al. 2003): [3.6], [4.5],
[5.8], and [8.0] $\mu m$. The extinction $A_{K_s}$ was calculated
from extinction maps for each star in the catalogue
by Skowron et al. (2019). According to these
authors, the error in the distance to the Cepheids
in their catalogue is $\sim$5\%. Skowron et al. (2019)
estimated the ages using the technique of Anderson
et al. (2016) by taking into account the spin period of
the stars and the metallicity index.

The distribution of Cepheids younger than 120 Myr
in projection onto the Galactic $XY$ plane is given in
Fig. 2. The dark dots in this figure mark 363 Cepheids
located no farther than 5 kpc from the Sun, where the
sample satisfies the completeness condition. Only the
Carina-Sagittarius arm segment is well represented
in the figure. From these 363 Cepheids we determined
the velocity perturbation amplitudes $f_R$ and $f_\theta$ based
on a spectral analysis of the residual stellar velocities.
The value of $\sigma_R=15$~km s$^{-1}$ for these stars can be
estimated from the error per unit weight obtained
when solving the kinematic equations in Bobylev and
Bajkova (2022d). A direct calculation based on 363
Cepheids with measured line-of-sight velocities gave
a close value, $\sigma_R=14.9$~km s$^{-1}$.

 \subsection*{DISCUSSION}
Table 2 gives the estimates of $\Omega_p$ by various methods. These results were obtained mostly from such young objects as OB stars, open star clusters
(OSCs), and Cepheids.

As is well known, the first estimates of $\Omega_p$ found by Lin et al. (1969) and Yuan (1969) provoked a debate (Marochnik et al. 1972) about the choice of the most
probable value, 13 or 23~km s$^{-1}$. However, choosing the true value of $\Omega_p$ is a no less acute problem even now. A discussion of this problem can be found, for example, in Palou\v{s} et al. (1977), Palou\v{s} (1980), Marochnik and Suchkov (1981), Pichardo et al. (2003), or Martos et al. (2004).

Surprisingly, but there are estimates with small $\Omega_p=12$~km s$^{-1}$ obtained by Eilers et al. (2020) and Vall\'ee (2021) from present-day data even now. However, we did not include these estimates in the table by deeming them exotic.

In the table we did not include the results of simulations (for example, Quillen and Minchev 2005; Chakrabarty 2007; Michtchenko et al. 2018) obtained with prespecified $\Omega_p$ by deeming these estimates indirect. The studies where the authors either did not decide on the mean value of $\Omega_p$ (Griv et al. 2017)
or found $\Omega_p$ separately from a particular spiral arm segment (Naoz and Shaviv 2007) were not included either. However, it can be already seen that the present-day results lie in a very wide range of $\Omega_p$: 18--32~km s$^{-1}$.

 \subsection*{Direct Method}
A simple relation follows from Eq. (3):
\begin{equation}
  \Omega_p=\frac{\theta-\theta_{\rm birth}}{t},
 \label{Omega-Lin}
\end{equation}
where $\theta$ is the current position of the star, $\theta_{\rm birth}$ is the position angle corresponding to the birthplace of the
star, and $t$ is the age of the star. In Table 2 this method
is designated as ``$\theta-\theta_{\rm birth}$''.

Following Dias and L\'epine (2005), this $Omega_p$ estimation
method is called the direct one. It is applied
in those cases where the space velocities of stars,
their individual ages, and the membership in a specific
spiral arm are known. Of course, the case in point are
young stars affected by the spiral density wave. To
determine the birthplaces of stars $\theta_{\rm birth}$, one usually
constructs their Galactic orbits in the past using an
appropriate model of the Galactic gravitational potential.

Using the direct method of analysis as applied to a sample of young OSCs, Naoz and Shaviv (2007) found $\Omega_p=20.0^{+1.7}_{-1.2}$~km s$^{-1}$ kpc$^{-1}$ for the
Perseus arm and $\Omega_p=28.9^{+1.3}_{-1.2}$~km s$^{-1}$ kpc$^{-1}$ for
the Local Arm. For two Carina-Sagittarius arms
segments these authors found two values: 
$\Omega_{p1}=16.5^{+1.2}_{-1.4}$~km s$^{-1}$ kpc$^{-1}$ and
$\Omega_{p2}=29.8^{+0.6}_{-1.3}$~km s$^{-1}$ kpc$^{-1}$.

Dias et al. (2019) analyzed the kinematics of $\sim$440 OSCs younger than 50 Myr belonging to the Perseus, Local, and Carina-Sagittarius spiral arm
segments. Data from the Gaia DR2 catalogue were
used to calculate the average distances and proper
motions of the clusters. Based on the direct method,
with the determination of the OSC birthplace, the
pattern speed and the corotation radius were estimated
to be $\Omega_p=28.2\pm2.1$~km s$^{-1}$ kpc$^{-1}$ and
$R_{\rm cor}=8.51\pm0.64$~kpc, respectively. For the adopted
$R_0=8.3$~kpc and $V_0 = 240$~km s$^{-1}$ the corotation
radius here is $R_{\rm cor}=(1.02\pm0.07) R_0$.

Based on various publications, Joshi and Malhotra (2023) produced a sample of 6133 OSCs most of which were detected already from Gaia data. Having analyzed the spatial distribution of these clusters, these authors showed that most of the OSCs left the spiral arms approximately 10--20 Myr after their formation. Having compared the
current positions of $\sim$440 young OSCs with their positions at birth, Joshi and Malhotra (2023) found $\Omega_p=26.5\pm1.5$~km s$^{-1}$  kpc$^{-1}$ based on relation (13). These authors estimated the corotation radius  $(R_{\rm cor}/R_0=1.08^{+0.06}_{-0.05})$ by tying to the
Galactic rotation curve specified by the potential
from Bovy (2015), where $\Omega_0=27.5$~km s$^{-1}$ kpc$^{-1}$
($V_0=220$~km s$^{-1}$ and $R_0=8.0$~kpc). Note that these
authors considered a more complex version of Eq.~(1)
describing each spiral arm segment in the form of a sector structure:
 \begin{equation}
 R=R_{\rm kink} e^{(\theta-\theta_{\rm kink})\tan i},
 \label{spiral-100}
 \end{equation}
where $R_{\rm kink}$ and $\theta_{\rm kink}$ are the characteristics of the
sector structure of the spiral arm.

%%%%%%%%%%%%%%%%%%%%%%%%%%%%%%%%%%%%%%%%%%%%%%%%%%%%%%%%%%%%% t-Oort
  \begin{table}[t]
  \caption[]{\small
$\Omega_p$ estimated by various authors from young objects
 }
  \begin{center}  \label{Table-2}    \small
  \begin{tabular}{|c|c|c|c|c|c|}\hline
  $\Omega_p,$~km s$^{-1}$ kpc$^{-1}$ & Objects & Method & Reference \\\hline

    $11-13.5$ & OB stars & $\theta-\theta_{\rm birth}$ &  (1) \\
    $\sim20$  & OB stars &                       $\nu$ &  (2) \\
 $17.8\pm3.1$ & OB stars &                       $\nu$ &  (3) \\
 $19.1\pm3.6$ & A,F,G supergiants and Cepheids &    $\nu$ &  (4) \\
       $21.3$ &      OSC & $\theta-\theta_{\rm birth}$ &  (5) \\
 $28.1\pm2.0$ & Cepheids &                       $\nu$ &  (6) \\
     $20\pm2$ &      OSC & $\theta-\theta_{\rm birth}$ &  (7) \\
    $\sim30$  & OB stars &                       $\nu$ &  (8) \\
    $24\pm1$  &      OSC & $\theta-\theta_{\rm birth}$ &  (9) \\
 $18.6^{+0.3}_{-0.2}$ & $\sim$200 000 RAVE stars  & $\nu$ & (10) \\
 $20.3\pm0.5$ &     OB stars & $\theta-\theta_{\rm birth}$ & (11) \\
 $23.0\pm0.5$ & OSCs and giants & $\theta-\theta_{\rm birth}$ & (12) \\
 $25.2\pm0.5$ &      Cepheids &                $\Delta\chi$ & (13) \\
 $28.2\pm2.1$ &           OSC & $\theta-\theta_{\rm birth}$ & (14) \\
 $32.0\pm0.9$ &           OSC & $\theta-\theta_{\rm birth}$ & (15) \\
     $27\pm1$ &      Cepheids &                $\Delta\chi$ & (16) \\
 $26.5\pm1.5$ &           OSC & $\theta-\theta_{\rm birth}$ & (17) \\

 \hline
 \end{tabular}\end{center}
 {\small
(1) Lin et al. (1969); (2) Cr\'ez\'e and Mennessier (1973); (3) Byl and Ovenden (1978); (4) Mishurov et al. (1979); (5) Loktin and Matkin (1992); (6) Mishurov et al. (1997); (7) Amaral and L\'epine (1997); (8) Fern\'andez et al. (2001); (9) Dias and L\'epine (2005); (10) Siebert et al. (2012); (11) Silva and Napiwotzki (2013); (12) Junqueira et al. (2015); (13) Dambis et al. (2015); (14) Dias et al. (2019); (15) Castro-Ginard et al. (2021); (16) Bobylev and Bajkova (2022e); (17) Joshi and Malhotra (2023).
 }
 \end{table}
%%%%%%%%%%%%%%%%%%%%%%%%%%%%%%%%%%%%%%%%%%%%%%%%%%%%%%%%%%%%%%%%% t2

 \subsection*{Relative Methods}
Below we will describe the results obtained by several methods. We call them relative, since they directly depend on the adopted angular velocity of
Galactic rotation $\Omega_0$. in Table 2 the method based on
the application of relations (6)--(9) is designated as ``$\nu$.''

When considering the relative shifts in the positions
of stars caused by the spiral density wave in a
time interval $\Delta t$, Eq. (3) will be written as
\begin{equation}
 \begin{array}{lll}
 \renewcommand{\arraystretch}{3.6}
 \displaystyle
  \ln\biggl(\frac{R}{R_0}\biggr)= \tan i
     \biggl[\theta-\theta_0+\frac{\chi-\chi_0}{m}+(\Omega-\Omega_p)\Delta t\biggr],
 \label{spiral-55}
 \end{array}
\end{equation}
At $R=R_0$ and $\theta=\theta_0=0$ we will have the following relation:
\begin{equation}
 \renewcommand{\arraystretch}{2.0}
  \Delta\Omega=\frac{\Delta\chi\times 10^3}{m \Delta t},
 \label{spiral-66}
\end{equation}
where the phase difference $\Delta \chi$ is in radians, the age
difference $\Delta t$ is in Myr, and $\Delta\Omega=\Omega_0-\Omega_p$ is in
km s$^{-1}$ kpc$^{-1}$. in Table 2 this method is designated as ``$\Delta\chi$''.

 \subsection*{Analysis of the Stellar Positions}
Having analyzed the spatial positions of OSCs with various ages, Loktin and Matkin (1992) estimated $\Omega_p=21.3$~km s$^{-1}$ kpc$^{-1}$. From the spatial distribution of classical Cepheids Dambis et al. (2015) found $\Omega_p=25.2\pm0.5$~km s$^{-1}$ kpc$^{-1}$ averaged over three spiral arm segments.

By studying the distribution of Cepheids with various ages in the Carina-Sagittarius and Outer arms, Bobylev and Bajkova (2022d) estimated $\Omega_p=27\pm1$~km s$^{-1}$ kpc$^{-1}$ and the corotation radius $R_{\rm cor}=9.0\pm0.3$~kpc $(R_{\rm cor}/R_0=1.1\pm0.04)$.

 \subsection*{Analysis of the Stellar Kinematics}
First note the results of applying relations (6)--(9). For example, Cr\'ez\'e and Mennessier (1973) found $\Omega_p=20.0\pm4.1$~km s$^{-1}$ kpc$^{-1}$ from a sample
of OB3 stars with the adopted $R_0=8$~kpc. Based on the kinematics of 183 AFG supergiants and a sample of 192 classical Cepheids, Mishurov et al. (1979) estimated
$\Omega_p=19.1\pm3.6$~km s$^{-1}$ kpc$^{-1}$. Later, based on the kinematics of classical Cepheids, Mishurov et al. (1997) found $\Omega_p=28.1\pm2.0$~km s$^{-1}$ kpc$^{-1}$. Fern\'andez et al. (2001) used this approach to study
the kinematics of OB stars from the Hipparcos catalogue
(1997) and obtained $\Omega_p$ close to 30~km s$^{-1}$ kpc$^{-1}$.

L\'epine et al. (2001) applied this approach to justify the model consisting of a superposition of two- and four-armed spiral patterns. In particular, based on a
sample of classical Cepheids with proper motions and line-of-sight velocities, they found $\Omega_p-\Omega_0=0.15$ and 0.18~km s$^{-1}$ kpc$^{-1}$ for the two- and four-armed spiral patterns, respectively. Thus, in this model the
Sun is virtually at the corotation radius, with the corotation circle being slightly closer to the Galactic center than the Sun. This follows from the fact that
the difference $\Omega_p-\Omega_0$ was found to be positive.

A positive difference $\Omega_p-\Omega_0\approx0.5$~km s$^{-1}$ kpc$^{-1}$
was also obtained, for example, by Mishurov and Zenina (1999). These authors analyzed a sample of classical Cepheids with proper motions from
Hipparcos (1997) and line-of-sight velocities and
found $\Omega_0=27.3\pm1.7$~km s$^{-1}$ kpc$^{-1}$ for $R_0=7.5\pm0.1$~kpc. As a result, they concluded that the Sun is close to the corotation circle, since the difference $R_{\rm cor}-R_0$ was $\approx$0.1 kpc.

In many of the cases listed in Table 2, where $\Omega_p$ 
was estimated by applying the $\Omega_p$  method, the difference
$\Omega_p-\Omega_0$ is negative and, therefore, $R_{\rm cor}>R_0$.

Based on 213 713 stars from the RAVE catalogue (Steinmetz et al. 2006), Siebert et al. (2012) estimated $\Omega_p=18.6^{+0.3}_{-0.2}$~km s$^{-1}$ kpc$^{-1}$ for $m=2$ using relations (6)--(9). Note that these authors also analyzed the four-armed model of the spiral structure ($m=4$) and found $\Omega_p=25.8^{+0.1}_{-0.1}$~km s$^{-1}$ kpc$^{-1}$.

The approach based on relation (16) is also applied. For example, Bobylev and Bajkova (2012) traced the change in radial phase $\Delta\chi$ obtained through a spectral analysis of the residual Cepheid velocities. As a result, from three samples of classical Cepheids with various ages, they found 
 $\Omega_0-\Omega_p=10\pm3$~km s$^{-1}$ kpc$^{-1}$ for the adopted $m=2$ (then,
 $\Omega_0-\Omega_p=5\pm2$~km s$^{-1}$ kpc$^{-1}$ for $m=4$). Thus,
taking $\Omega_0=29$~km s$^{-1}$ kpc$^{-1}$ for Cepheids, we
estimate $\Omega_p=24\pm2$~km s$^{-1}$ kpc$^{-1}$ for $m=4$, which
is in good agreement with the results in Table~1.

 \section*{CONCLUSIONS}
In this paper we used the estimates of the amplitudes
of the velocity perturbations $f_R$ and $f_\theta$ caused
by the influence of a spiral density wave that were obtained
by us previously from various stellar samples.
These included: (i) Galactic masers with measured
VLBI trigonometric parallaxes and proper motions,
(ii) OB2 stars, and (iii) Cepheids. The proper motions
of the OB2 stars and Cepheids were taken from the
Gaia EDR3 catalogue.

The distances to the masers used were measured
with errors less than 10\%. The errors in the distances
to the OB2 stars that were calculated based
on their trigonometric parallaxes from the Gaia EDR3
catalogue have the same level. The distances to the
Cepheids were calculated by Skowron et al. (2019)
based on the period-luminosity relation with errors
less than 5\%. For all three samples the velocity
perturbations $f_R$ and $f_\theta$ were found using a spectral
analysis.

From these data we obtained new estimates of the
spiral pattern speed 
 $\Omega_p:$ $24.61\pm2.06$, $24.71\pm1.29$ and $25.98\pm1.37$~km s$^{-1}$ kpc$^{-1}$ from the samples of masers, OB2 stars, and Cepheids, respectively. The
corotation radii $R_{\rm cor}/R_0$ for these three samples are
$1.16\pm0.09$, $1.15\pm0.06$ and $1.09\pm0.06$, suggesting
that corotation is close to the Sun, with the corotation circle being located between the Sun and the Perseus arm segment.

The results obtained by us from these three samples are in excellent agreement between themselves. However, the $\Omega_p$ estimates obtained by other authors in recent years lie in a fairly wide range of $\Omega_p: 18-32$~km s$^{-1}$ kpc$^{-1}$.

 \subsubsection*{ACKNOWLEDGMENTS}
We are grateful to the referee for the useful remarks that
contributed to an improvement of the paper.

 \subsubsection*{REFERENCES}
 \small

\quad\quad 1. I. A. Acharova, J. R. D. L\'epine, Yu. N. Mishurov, B.M. Shustov, A. V. Tutukov, and D. S. Wiebe, Mon. Not. R. Astron. Soc. 402, 1149 (2010).

2. L. H. Amaral and J. R. D. L\'epine, Mon. Not. R. Astron. Soc. 286, 885 (1997).

3. R. I. Anderson, H. Saio, S. Ekstr\"om, C. Georgy, and
G. Meynet, Astron. Astrophys. 591, A8 (2016).

4. R. A. Benjamin, E. Churchwell, B. L. Babler, L. Brian,
T. M. Bania, D. P. Clemens, M. Cohen, J. M. Dickey,
et al., Publ. Astron. Soc. Pacif. 115, 953 (2003).

5. V. V. Bobylev and A. T. Bajkova, Astron. Lett. 38, 638 (2012).

6. V. V. Bobylev and A. T. Bajkova, Mon. Not. R. Astron. Soc. 437, 1549 (2014).

7. V. V. Bobylev and A. T. Bajkova, Astron. Rep. 65, 498 (2021).

8. V. V. Bobylev and A. T. Bajkova, Astron. Rep. 66, 269 (2022) a.

9. V. V. Bobylev and A. T. Bajkova, Astron. Lett. 48, 376 (2022) b.

10. V. V. Bobylev and A. T. Bajkova, Astron. Lett. 48, 169 (2022) c.

11. V. V. Bobylev, A. T. Bajkova, Astron. Rep. 66, 545 (2022) d.

12. V. V. Bobylev and A. T. Bajkova, Astron. Lett. 48, 568 (2022) e.

13. J. Bovy, Astrophys. J. Suppl. Ser. 216, 29 (2015).

14. J. Byl and M. W. Ovenden, Astrophys. J. 225, 496 (1978).

15. A. Castro-Ginard, P. J. McMillan, X. Luri, 
 %C. Jordi, M. Romero-Gomez, T. Cantat-Gaudin, L. Casamiquela, Y. Tarricq, C. %Soubiran, and F. Anders, 
et al., Astron. Astrophys. 652, 162 (2021).

16. D. Chakrabarty, Astron. Astrophys. 467, 145 (2007).

17. X. Chen, S.Wang, L. Deng, R. de Grijs, and M. Yang,
Astrophys. J. Suppl. Ser. 237, 28 (2018).

18. M. Cr\'ez\'e and M. O. Mennessier, Astron. Astrophys. 27, 281 (1973).

19. A. K. Dambis, L. N. Berdnikov, Yu. N. Efremov, A.
 %Yu. Kniazev, A. S. Rastorguev, E. V. Glushkova, V. V.
 %Kravtsov, D. G. Turner, D. J.Majaess, and R. Sefako,
et al., Astron. Lett. 41, 489 (2015).

20. W. S.Dias and J. R. D. L\'epine, Astrophys. J. 629, 825 (2005).

21. W. S. Dias, H. Monteiro, J. R. D. L\'epine, and D. A. Barros, Mon. Not. R. Astron. Soc. 486, 5726 (2019).

22. A.-C. Eilers, D. W. Hogg, H.-W. Rix, et al., Astrophys. J. 900, 186 (2020).

23. D. Fern\'andez, F. Figueras, and J. Torra, Astron. Astrophys.
372, 833 (2001).

24. O. Gerhard, Mem. Soc. Astron. It. Suppl. 18, 185 (2011).

25. E. Griv, L.-G. Hou, I.-G. Jiang, and C.-C. Ngeow,
Mon. Not. R. Astron. Soc. 464, 4495 (2017).

26. The HIPPARCOS and Tycho Catalogues, ESA SP--1200 (1997).

27. T. Hirota, T. Nagayama, M. Honma, Y. Adachi,
R. A. Burns, J. O. Chibueze, Y. K. Choi, K.
Hachisuka, et al. (VERA Collab.), Publ. Astron. Soc.
Jpn. 70, 51 (2020).

28. Y. C. Joshi and S. Malhotra, arXiv: 2212.09384 (2023).

29. T. C. Junqueira, C. Chiappini, J. R. D. L\'epine, et al.,
Mon. Not. R. Astron. Soc. 449, 2336 (2015).

30. J. R. D. L\'epine, Yu. N.Mishurov, and S. Yu. Dedikov,
Astrophys. J. 546, 234 (2001).

31. C. C. Lin and F. H. Shu, Astrophys. J. 140, 646 (1964).

32. C. C. Lin, C. Yuan, and F. H. Shu, Astrophys. J. 155, 721 (1969).

33. A. V. Loktin and N. V. Matkin, Astron. Astrophys. Trans. 3, 169 (1992).

34. L. S. Marochnik, Yu. N. Mishurov, and A. A. Suchkov, Astrophys. Space Sci. 19, 285 (1972).

35. L. S. Marochnik and A. A. Suchkov, Astrophys. Space Sci. 79, 337 (1981).

36. M. Martos, X. Hernandez,  M. Y\'a\~nez, E. Moreno, and
B. Pichardo, Mon. Not. R. Astron. Soc. 350, L47 (2004).

37. T. A. Michtchenko, J. R. D. L\'epine, A. P\'erez-Villegas, R. S. S. Vieira, and D. A. Barros, Astrophys. J. Lett. 863, L37 (2018).

38. Yu. N. Mishurov, E. D. Pavlovskaia, and A. A. Suchkov, Astron. Rep. 56, 268 (1979).

39. Yu. N. Mishurov, I. A. Zenina, A. K. Dambis, A. M. Mel'nik, and A. S. Rastorguev, Astron. Astrophys. 323, 775 (1997).

40. Yu. N.Mishurov and I. A. Zenina, Astron. Astrophys. 341, 781 (1999).

41. Yu. N. Mishurov and I. A. Acharova, Mon. Not. R. Astron. Soc. 412, 1771 (2011).

42. S. Naoz and N. J. Shaviv, New Astron. 12, 410 (2007).

43. J. Palou\v{s}, J. Ruprecht, O. B. Dluzhnevskaya, and
T. Piskunov, Astron. Astrophys. 61, 27 (1977).

44. J. Palou\v{s}, Astron. Astrophys. 87, 361 (1980).

45. B. Pichardo,M.Martos, E.Moreno, and J. Espresate, Astrophys. J. 582, 230 (2003).

46. A. C. Quillen and I. Minchev, Astron. J. 130, 576 (2005).

47. A. S. Rastorguev, E. V. Glushkova, M. V. Zabolotskikh,
and H. Baumgardt, Astron. Astrophys. Trans. 20, 103 (2001).

48. M. J. Reid, K. M. Menten, A. Brunthaler,
X.W. Zheng, T.M.Dame, Y. Xu, Y.Wu, B. Zhang, et
al., Astrophys. J. 783, 130 (2014).

49. M. J. Reid, K. M. Menten, A. Brunthaler,
X.W. Zheng, T.M. Dame, Y. Xu, J. Li, N. Sakai, et
al., Astrophys. J. 885, 131 (2019).

50. K. Rohlfs, Lectures on Density Wave Theory (Springer, Berlin, 1977).

51. A. Siebert, B. Famaey, J. Binney, 
 %B. Burnett, C. Faure, I. Minchev, M. E. K. Williams, O. Bienayme,
et al., Mon. Not. R. Astron. Soc. 425, 2335 (2012).

52. M. D. V. Silva and R. Napiwotzki, Mon. Not. R. Astron. Soc. 431, 502 (2013).

53. D. M. Skowron, J. Skowron, P. Mr\'oz, A. Udalski,
 %P. Pietrukowicz, I. Soszynski, M. K. Szymanski, R. Poleski, 
et al., Science (Washington, DC, U. S.) 365, 478 (2019).

54. M. Steinmetz, T. Zwitter, A. Siebert, F. G. Watson,
 %K. C. Freeman, U. Munari, R. Campbell, M. Williams, 
et al., Astron. J. 132, 1645 (2006).

55. A. Udalski, M. K. Szyma\'nski, and G. Szyma\'nski,
Acta Astron. 65, 1 (2015).

56. J. P. Vall\'ee, Astrophys. J. 454, 119 (1995).

57. J. P. Vall\'ee, Astrophys. J. 566, 261 (2002).

58. J. P. Valle\'e, Astron. J. 135, 1301 (2008).

59. J. P. Vall\'ee, New Astron. Rev. 79, 49 (2017a).

60. J. P. Vall\'ee, Astrophys. Space Sci. 362, 79 (2017b).

61. J. P. Vall\'ee, Mon. Not. R. Astron. Soc. 506, 523 (2021).

62. S. Wang, X. Chen, R. de Grijs, and L. Deng, Astrophys. J. 852, 78 (2018).

63. Y. Xu, L. G. Hou, S. Bian, et al., Astron. Astrophys. 645, L8 (2021).

64. C. Yuan, Astrophys. J. 158, 889 (1969).

 \end{document}